\documentclass{ws-mpla}
\usepackage{graphicx}
\usepackage{amssymb}

\begin{document}

\title{EMPIRICAL SURVEY OF NEUTRINOLESS DOUBLE BETA DECAY MATRIX ELEMENTS}
\author{\footnotesize R.\,G.\,H. ROBERTSON}
\address{Dept. of Physics and Center for Experimental Nuclear Physics and Astrophysics \\ University of Washington, Seattle, WA 98195, USA\\
rghr@uw.edu \\
}

\maketitle

\pub{Received ()}{}

\begin{abstract}
Neutrinoless double beta decay has been the subject of intensive theoretical work as it represents the only practical approach to discovering whether neutrinos are Majorana particles or not, and whether lepton number is a conserved quantum number.  Available calculations of matrix elements and phase-space factors are reviewed from the perspective of a future large-scale experimental search for $0\nu\beta\beta$ decay.  Somewhat unexpectedly, a uniform inverse correlation between phase space and the square of the nuclear matrix element emerges.  As a consequence, no isotope is either favored or disfavored; all have qualitatively the same decay rate per unit mass for any given value of the Majorana mass.

\end{abstract}

\ccode{PACS Nos.: 14.60.Pq, 23.40.Hc }

\section{Introduction}

The matter asymmetry of the universe remains one of the deepest mysteries in physics.  The 
absence of significant amounts of antimatter requires, as Sakharov explained,\cite{Sakharov:1967dj} a time when 
the universe was not in equilibrium, the non-conservation of baryon  number, and 
violation of CP invariance.  Non-conservation of baryon and lepton number has not been 
experimentally discovered, despite heroic efforts to observe proton decay.    The observation of neutrinoless double beta decay would  demonstrate the  non-conservation of lepton number, and by inference the non-conservation of baryon number.  It is the only known practical approach to discovering whether or not neutrinos are their own antiparticles, i.e. `Majorana' particles.

Neutrinoless double beta decay can be induced by the exchange of a massive Majorana neutrino with an electron-flavor admixture.  From neutrino-oscillation data neutrinos are known to have mass, and are expected on general theoretical grounds to have a Majorana character.  The ``see-saw'' mechanism proposed  to explain the lightness of neutrinos (see Ref. \refcite{Lindner:2007rs} and references contained therein) predicts that they are Majorana particles.  The effective Majorana mass may be complicated by other kinds of new physics and might not be the physical mass of a particle, but it serves as a  metric for designing and comparing experiments.  

The rate of neutrinoless double beta decay ($0\nu\beta\beta$) depends on the available phase space, the size of the nuclear matrix element, and the effective Majorana neutrino mass.\cite{Elliott:2012sp,Vergados:2012xy} Specifically, the half-life $\tau_{1/2}^{0\nu}$ is given by
\begin{eqnarray}
\left[\tau_{1/2}^{0\nu}\right]^{-1} &=&G_{0\nu}^{(0)}g_A^4\left|M_{0\nu}\right|^2\left|\frac{\langle m_{ee} \rangle}{m_e}\right|^2
\end{eqnarray}
where $G_{0\nu}^{(0)}$ is the phase space factor as defined and tabulated by Kotila and Iachello\cite{PRD23_649_1981}, $M_{0\nu}$ is the nuclear matrix element, and the Majorana mass is:
\begin{eqnarray}
\langle m_{ee}\rangle &=& \left|U_{e1}^2m_1+U_{e2}^2m_2e^{i\alpha}+U_{e3}^2m_3e^{i\beta}\right|.
\end{eqnarray}
The convention for the Majorana phases is the one given by Rodejohann.\cite{Rodejohann:2011mu}   The Majorana mass is a coherent sum over mass eigenstates with (potentially) CP-violating phases, and cancellations can occur.  The effective mass can also be modified by interference with other hypothesized non-standard-model processes.

Historically the effective axial-vector coupling constant $g_A$ has usually  been incorporated in the phase space factor, and sometimes in the nuclear matrix element.  Here, following Kotila and Iachello, we break it out explicitly.  The phase space factor $G_{0\nu}^{(0)}$ has recently been reevaluated by Kotila and Iachello (Table III in Ref. \refcite{PRD23_649_1981})  with an exact treatment of screening, resulting in significant downward corrections of as much as a factor of 2 for the heaviest nuclei.   These authors use for the nuclear radius $R=r_0A^{1/3}$ with $r_0=1.2$ fm, cautioning that some authors have used $r_0=1.1$ fm.

The presence of $g_A^4$ in the rate introduces a significant uncertainty in the calculated rates, in addition to the well-known uncertainty in the nuclear matrix element.  Barea {\em et al.}\cite{PRL109_042501_2012} and Ejiri\cite{PPNP64_249_2010} have fitted the known half-lives for $2\nu\beta\beta$ decay and find effective values of $g_A{\rm} $ of about 0.8 for shell-model calculations and 0.6 for the Interacting Boson Model (IBM).  Barea {\em et al.}\cite{PRL109_042501_2012}  also note a weak A-dependence, which we neglect here.  With these renormalized values there is qualitatively good agreement with the data on $2\nu\beta\beta$ half-lives.  In contrast, the calculated phase-space factors for neutrinoless decay are generally presented with the free-nucleon value $g_A=1.269$, $g_A=1.25$, or $g_A=1$.  The difference between the free-nucleon value for $g_A$ and 0.6 corresponds to a factor of 20 in rate.  The extent of the renormalization of $g_A$ in neutrinoless double beta decay remains a topic of discussion among theorists.


\section{Application of Theory to Experiments}

Experimental work on at least nine different double beta unstable
nuclides is in progress around the world, not counting the more unconventional projects involving electron capture and positron emission.  Experiments using large amounts of $^{76}$Ge, $^{130}$Te, $^{136}$Xe, $^{150}$Nd 
are actively being prepared or planned by international teams involving US researchers.  The Super-NEMO Collaboration is focusing its attention on a large $^{82}$Se experiment.\cite{Barabash:2012hn}
A fundamental requirement is that, independent of technical issues concerning background, resolution, etc., there must be sufficient signal to detect.   The sensitivity to a particular value of the Majorana mass depends on the phase space for the decay, the nuclear matrix element, the effective value in the nuclear medium for the axial-vector coupling constant $g_A$, the equivalent mass of isotope $M$ (containing $N$ atoms) that is actively monitored at 100\% efficiency, and the time over which the measurement is made.  The decay rate per unit mass of isotope is
\begin{eqnarray}
\lambda_{0\nu}\frac{N}{M} &=& \frac{\ln(2)N_A}{A m_e^2}G_{0\nu}^{(0)}g_A^4\left|M_{0\nu}\right|^2\left| \langle m_{ee} \rangle \right|^2 \nonumber \\
&\equiv & H_{0\nu}g_A^4\left|M_{0\nu}\right|^2\left| \langle m_{ee} \rangle \right|^2 \label{eq:rate}
\end{eqnarray}
where $A$ is the atomic mass of the isotope,  and $N_A$ is Avogadro's number.  Constants are aggregated with $G_{0\nu}^{(0)}$ to form the `specific phase space' $H_{0\nu}$.   The phase space $G_{0\nu}^{(0)}$ is an activity per atom, whereas the specific phase space $H_{0\nu}$ is an activity per unit mass.  The nuclear matrix elements have been calculated by a number of methods, and we make use of a recent compilation by Dueck {\em et al.}\cite{PRD83_113010_2011} supplemented with results for $^{128}$Te in Refs. \refcite{PRC79_055501_2009,Faessler:2008xj} (for a still more recent evaluation, see Ref. \refcite{BareaPRCinpress}).
\begin{table}[h]
\tbl{Phase-space factors 
$G_{0\nu}^{(0)}$ in units of $\times 10^{-15}$ y$^{-1}$
from ref.~\protect\refcite{PRD23_649_1981}, 
specific phase space $H_{0\nu}$ in units of Mg$^{-1}$ y$^{-1}$   eV$^{-2}$ from Eq.~\ref{eq:rate}, and nuclear matrix elements from refs.~\protect\refcite{PRD83_113010_2011,Iachello:2011zzd} 
for neutrinoless double beta decay candidate isotopes.}
{\begin{tabular}{lrrrrrrrrrrr} 
\toprule
 Isotope & Q &$G_{0\nu}^{(0)}$ & $H_{0\nu}$ & Shell  & GCM &QRPA  & QRPA  & IBM  & IBM & PHFB & PHFB    \\
 & keV & & & Model&&low&high&low&high&low&high \\
  \colrule
$^{48}$Ca&4272&24.81&826.2&0.85&2.37&&&&2.00&&\\
$^{76}$Ge&2039&2.36&49.6&2.81&4.60&4.20&7.24&4.64&5.47&&\\
$^{82}$Se&2995&10.16&198.1&2.64&4.22&2.94&6.46&3.81&4.41&&\\
$^{96}$Zr&3350&20.58&342.7&&5.65&1.56&3.12&&2.53&2.24&3.46\\
$^{100}$Mo&3034&15.92&254.5&&5.08&3.10&6.07&3.73&4.22&4.71&7.77\\
$^{110}$Pd&2018&4.82&70.0&&&&&&3.62&5.33&8.91\\
$^{116}$Cd&2814&16.70&230.1&&4.72&2.51&4.52&&2.78&&\\
$^{124}$Sn&2287&9.04&116.5&2.62&4.81&&&&3.53&&\\
$^{128}$Te&866&0.59&7.4&&4.11&3.50&6.16&&4.52&&\\
$^{130}$Te&2527&14.22&174.8&2.65&5.13&3.19&5.50&3.37&4.06&2.99&5.12\\
$^{136}$Xe&2458&14.58&171.4&2.19&4.20&1.71&3.53&&3.35&&\\
$^{148}$Nd&1929&10.10&109.1&&&&&&1.98&&\\
$^{150}$Nd&3371&63.03&671.7&&1.71&3.45&&2.32&2.89&1.98&3.70\\
$^{154}$Sm&1215&3.02&31.3&&&&&&2.51&&\\
$^{160}$Gd&1730&9.56&95.5&&&&&&3.63&&\\
$^{198}$Pt&1047&7.56&61.0&&&&&&1.88&&\\
\botrule
\end{tabular}}
\label{tab:me}
\end{table}
The matrix elements compiled by Dueck {\em et al.} and shown in Table \ref{tab:me} were renormalized by them to a common value $g_A = 1.25$.  We renormalize again to remove $g_A$ from the matrix element entirely, i.e.  divide the tabulated numbers by $1.25^2$ and square to obtain the quantity $\left|M_{0\nu}\right|^2$ of Eq. \ref{eq:rate}.  

The possible range for the effective value of $g_A$ in each case is taken to be not larger than the free nucleon value, 1.269, and not smaller than 0.8 for shell-model matrix elements or 0.6 for other calculation methods (GCM, QRPA,  IBM, and PHFB).  The lower `limits' are guided by the results reported by  Barea {\em et al.}\cite{PRL109_042501_2012,BareaPRCinpress} and  based on the renormalization needed to fit experimental lifetimes for $2\nu\beta\beta$ decays.  However, there is no consensus that such a large renormalization will apply to $0\nu\beta\beta$ decays.  It can be argued that in $0\nu\beta\beta$ a closure approximation is applicable, and all intermediate states and multipoles are included, whereas in $2\nu\beta\beta$ the strength is concentrated in just a few intermediate 1$^+$  states.  Quenching of $g_A$ is a symptom of basis truncation, and it might therefore be less of an issue for $0\nu\beta\beta$.  Furthermore, many calculations make empirical adjustments to internal parameters such as $g_{pp}$, the particle-particle strength in QRPA, or occupation numbers determined from two-nucleon transfer reactions.  A detailed discussion is given by Faessler {\em et al.}\cite{Faessler:2008xj}  These adjustments have the effect of correcting for some of the basis truncation, a correction that may also diminish the need to renormalize $g_A$.

\begin{figure}

\centering

\includegraphics[height=2.7in]{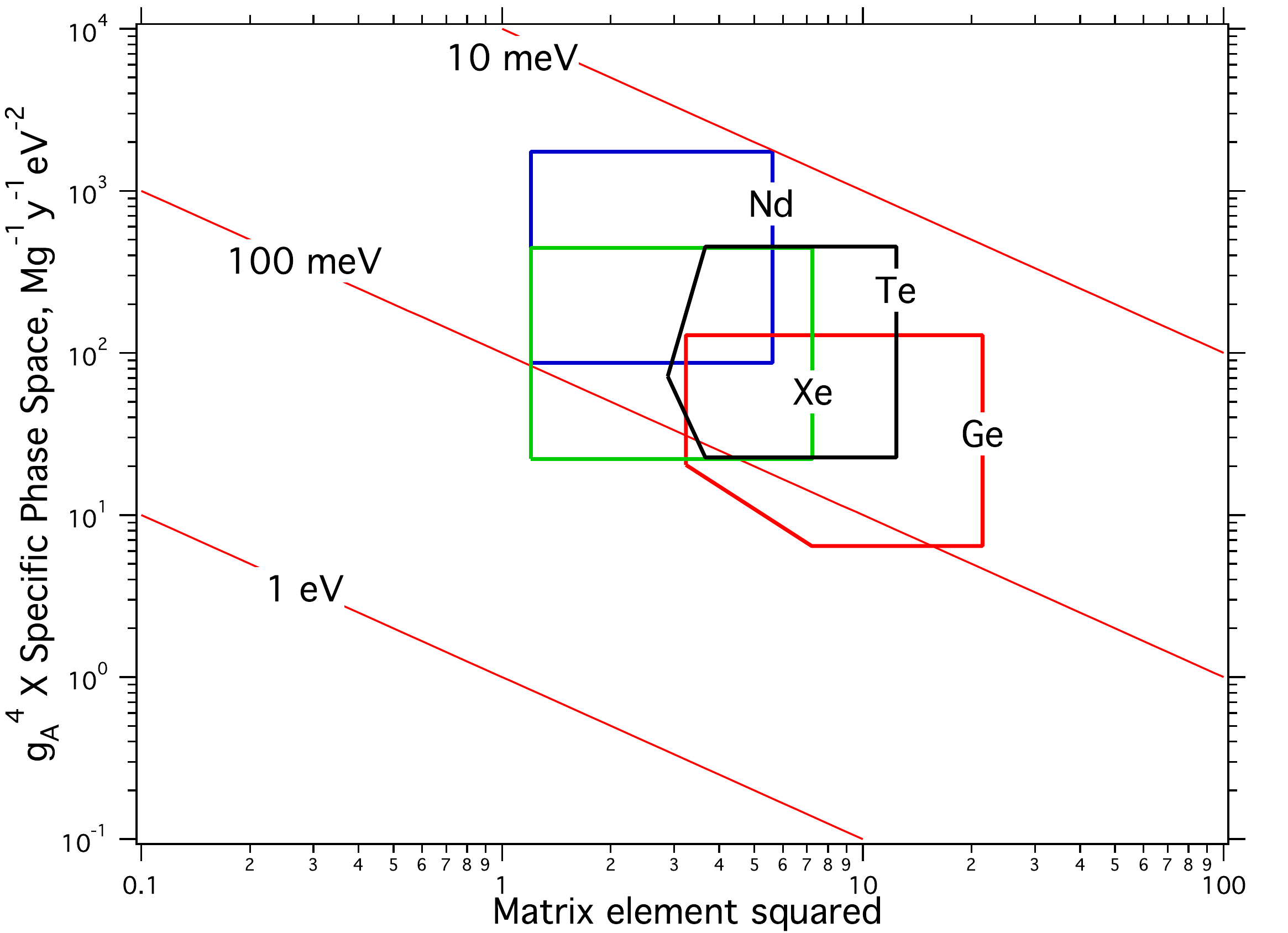}

\caption{Regions in the renormalized specific phase space  $g_A^4 H_{0\nu} = g_A^4\ln(2)\frac{N_A}{Am_e^2}G_{0\nu}^{(0)}$ and matrix element squared $\left|M_{0\nu}\right|^2$ that encompass modern theoretical calculations, for the candidate neutrinoless double beta decay isotopes $^{76}$Ge, $^{130}$Te, $^{136}$Xe, and $^{150}$Nd.  The vertical span reflects the range of $g_A$, which differs for the shell model and other models, leading to non-rectangular boundaries.  The matrix-element calculational methods are shell model (SM), generator-coordinate method (GCM), quasiparticle random-phase approximation (QRPA),  interacting boson model (IBM), and Projected Hartree-Fock Bogoliubov method (PHFB), as given in Table \ref{tab:me}. The lines indicate the effective Majorana mass that would correspond to a count rate of 1 event per tonne per year.}

\label{fig:DBDrate}

\end{figure}

While the coupling constant $g_A^2$ is formally a factor in the nuclear matrix element, it has been given an independent role in theoretical calculations.   The question of how $g_A$ is renormalized is decoupled from other sources of uncertainty, such as the initial- and final-state wave functions and short-range correlations.   To keep these two classes of uncertainty distinct, in the following two plots we multiply the specific phase space by $g_A^4$ and plot the product on the vertical axis against the square of the nuclear matrix element on the horizontal axis.  Since the specific phase space itself is essentially exact, this provides a convenient way of displaying both sources of uncertainty, and correlations between them can be included on a theory-by-theory basis as mentioned.  Figure \ref{fig:DBDrate} summarizes the rates and theoretical ranges for  four important isotopes.    It may be seen that the four isotopes have comparable sensitivity.
The vertical span of the regions results from the spread in $g_A$ and  in a Bayesian sense may be thought of as an uncertainty.  The true value is likely to be found in this range.  For the horizontal span, an interpretation as an uncertainty is much less satisfactory.  Different theoretical approaches do explore different types of deficiency in theory, there being no exact theory for this process, but they do not necessarily include the true value.   Moreover, it can be expected that better, or at least different, theories applied in the future might expand the ranges shown: no mechanism can reduce a range that encompasses all values.   Such behavior is in contrast to what one expects for an uncertainty.  Despite these shortcomings, there is comparative information available at a glance about the theoretical situation for each isotope in these plots. 

\begin{figure}

\centering

\includegraphics[height=2.7in]{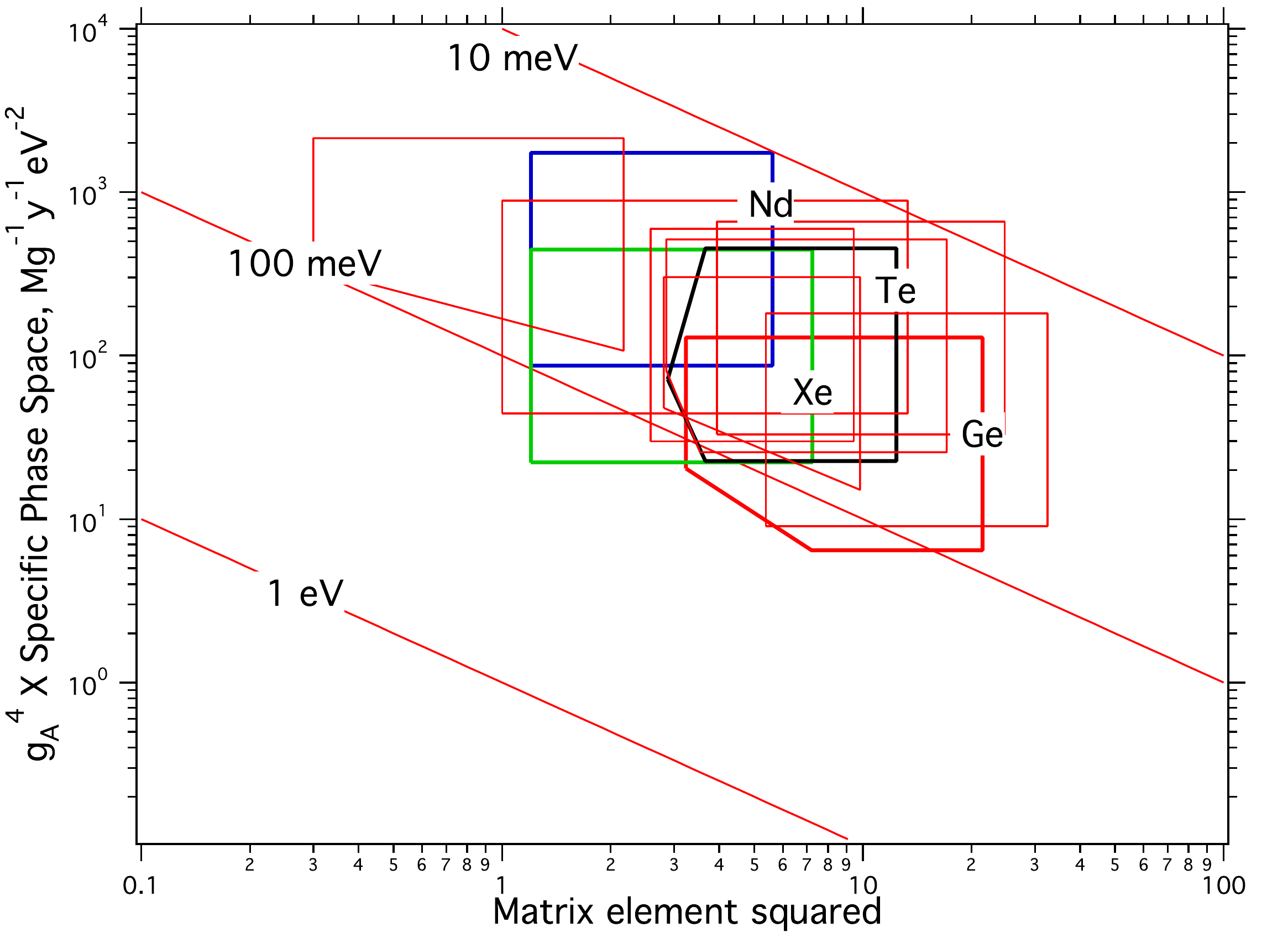}

\caption{As Fig.~\ref{fig:DBDrate} but with the addition of the isotopes $^{48}$Ca (2.2, 2143), $^{82}$Se (17, 514), $^{96}$Zr (13, 889), $^{100}$Mo (25, 660), $^{110}$Pd (33,181), $^{116}$Cd (9, 597), and $^{124}$Sn (10, 302).  The number pairs are the coordinates of the upper rightmost corner of each area, in lieu of labeling.  It is more difficult to see the details but the overall trend of a correlation between the phase space factor and the square of the nuclear matrix element is brought out.}

\label{fig:DBDrateAll}

\end{figure}
\begin{figure}

\centering

\includegraphics[height=2.7in]{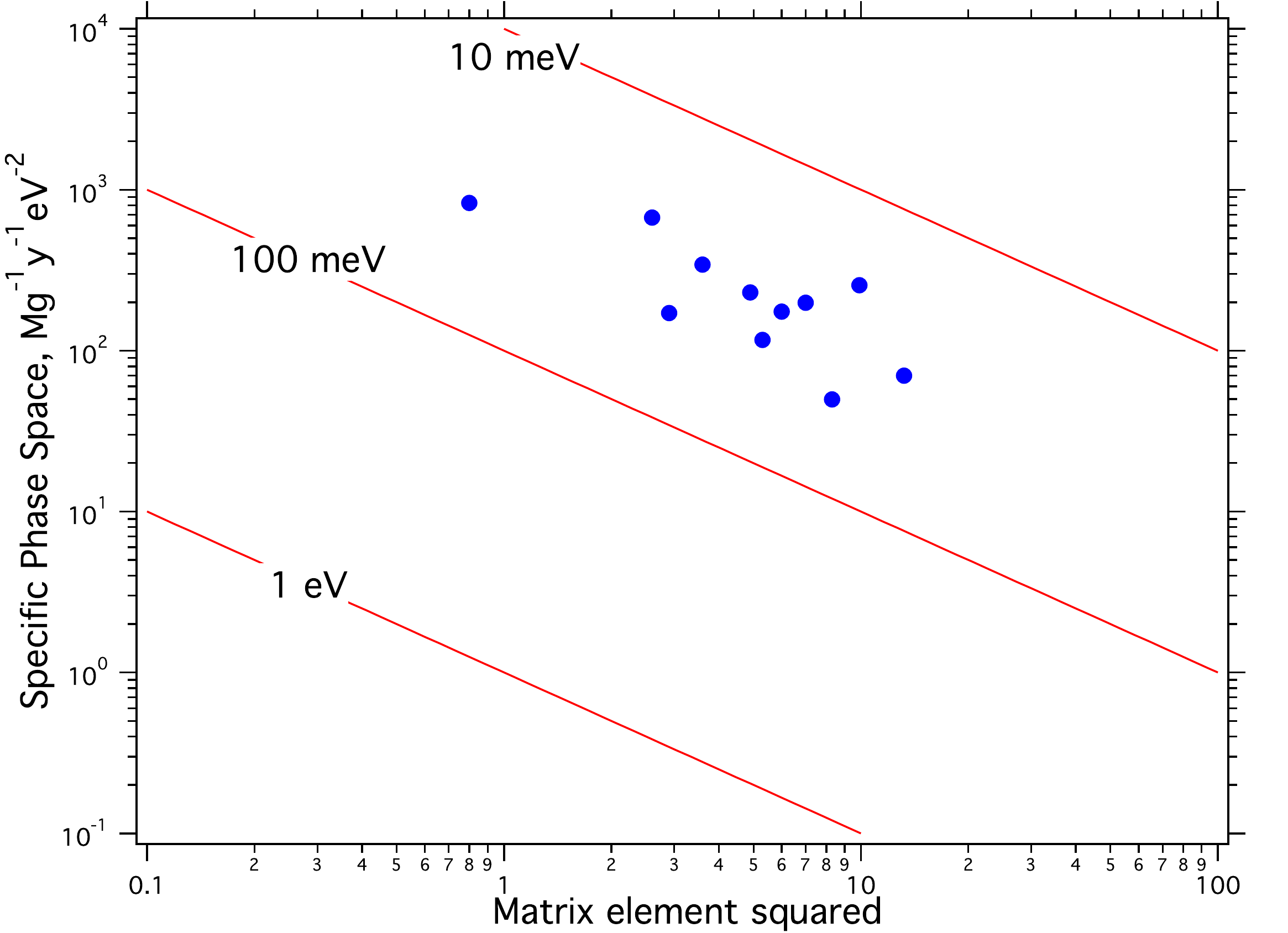}

\caption{For each candidate isotope a point is plotted at the geometric mean of the squared matrix element range limits (as shown in Fig.~\ref{fig:DBDrateAll}) and the phase-space factor evaluated at $g_A$=1.  The points in order of increasing abscissa value are:   $^{48}$Ca, $^{150}$Nd, $^{136}$Xe,  $^{96}$Zr, $^{116}$Cd, $^{124}$Sn, $^{130}$Te,  $^{82}$Se, $^{76}$Ge, $^{100}$Mo, and $^{110}$Pd.}

\label{fig:DBDrateMean}

\end{figure}

For clarity we present the results in two plots, the second plot, Fig.~\ref{fig:DBDrateAll}, containing the same isotopes as Fig.~\ref{fig:DBDrate} but in addition the isotopes $^{48}$Ca, $^{82}$Se, $^{96}$Zr, $^{100}$Mo, $^{110}$Pd, $^{116}$Cd, and $^{124}$Sn,  confining our attention to Q-values $\geq 2$ MeV.   
In this figure a correlation between the phase-space factor and the nuclear matrix element is strikingly apparent, and there do not even seem to be significant exceptions.  This seems quite surprising, and to our knowledge has not previously been  remarked on.    The trends in these plots are linear in a log-log display but would be hyperbolae and not easily seen in linear-scale plots. 
 A different view with the same information is shown in Fig.~\ref{fig:DBDrateMean}.  Here the geometric means of the squared matrix elements are plotted for a phase-space factor with $g_A=1$.  
The nucleus $^{48}$Ca is generally considered to have a `hindered' $0\nu\beta\beta$ matrix element, but it does not appear to be in any sense unusual  in these figures.   One general conclusion is that (at the level of a factor of 3 or so), there are no especially favored or disfavored isotopes for a $0\nu\beta\beta$ search.  They all have roughly equivalent sensitivity from the theoretical standpoint.  The specific activity may be expressed  as
\begin{eqnarray}
\lambda_{0\nu}\frac{N}{M} &=& a_{0\nu} g_A^4 \left| \langle m_{ee} \rangle \right|^2  \label{eq:powerlaw} \\
\log\left(\lambda_{0\nu}\frac{N}{M}\right) &=& \log a_{0\nu} + 4\log g_A +2\log \left| \langle m_{ee} \rangle \right|, \nonumber
\end{eqnarray}
where $a_{0\nu} = \langle H_{0\nu}|M_{0\nu}|^2\rangle$ is a constant with a value $\sim 10^{2.9\pm0.5}$ decays  per year per tonne per eV$^2$.
Of course, for experiments that are potentially very costly, factors of a few are important, and the deviations from this general trend require the best theoretical treatment possible.

\begin{figure}

\centering

\includegraphics[height=2.7in]{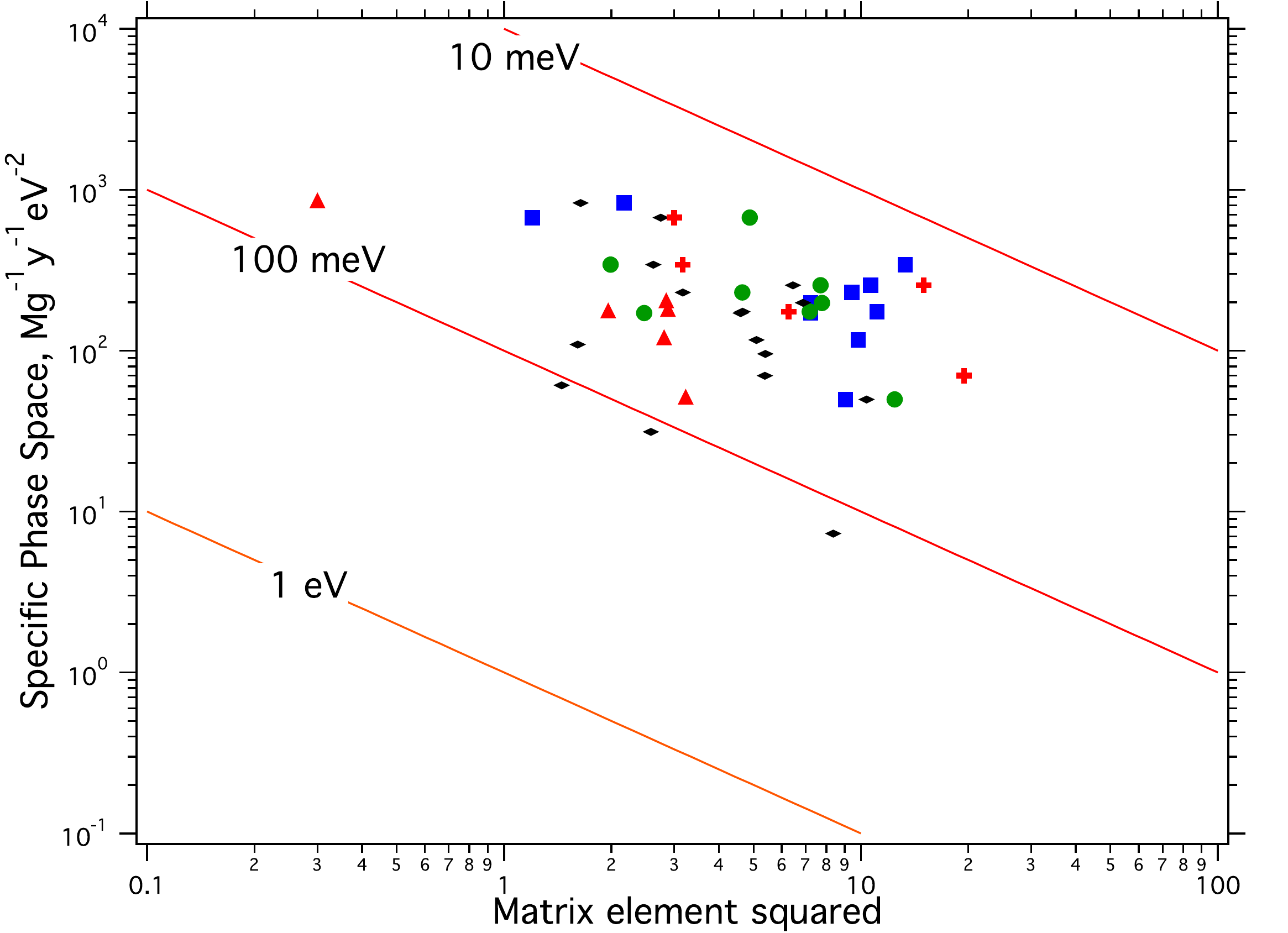}

\caption{The matrix elements determined individually theory by theory and listed in Table \ref{tab:me} are plotted against the specific phase space.  Where a theory predicts a range of values, the geometric means of the high and low predictions are plotted.  Red triangles: shell model, green circles: GCM, blue squares: QRPA, black diamonds: IBM, red crosses: PHFB. }

\label{fig:indiv}

\end{figure}

Why would a large phase-space factor imply a small matrix element and conversely?  For any given choice of $g_A$, the specific phase space as defined above depends explicitly on $A$, and implicitly on the Q-value and $Z$.   The presence of the factor $A$ in the specific phase space has little effect on the correlation but slightly reduces the scatter.    Most of the correlation therefore must be traceable to the Q-value and $Z$.   A natural question is whether individual theoretical predictions show the same trends as do the summary data.  The points are shown in Fig.~\ref{fig:indiv} and include isotopes with $Q \leq 2$ MeV.  The trend is less apparent, with the scatter of values from each theory dominating the distribution and the low-Q cases falling below the rest.  That is to be expected because a power-law scaling relation can only apply in a limited range of Q since, while the phase space factor can be arbitrarily small, the matrix element cannot increase without limit.   

\begin{figure}

\centering

\includegraphics[height=2.7in]{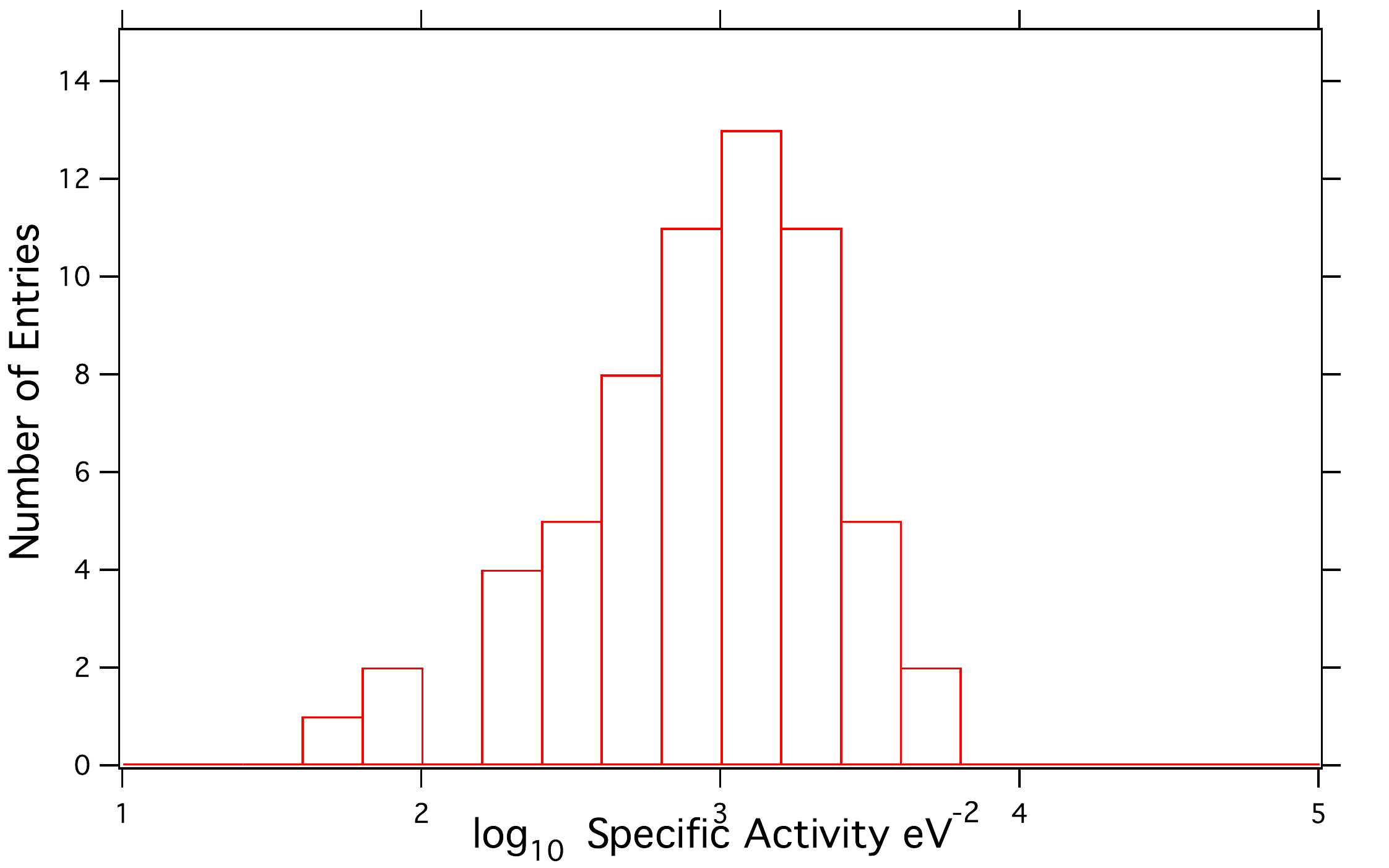}

\caption{The individual values of the specific activity determined by theory and listed in Table \ref{tab:me} are histogrammed for $g_A=1$.  No averaging is applied in this histogram; for theories with high and low entries, both are included. }

\label{fig:Bqhist}

\end{figure}

In Fig.~\ref{fig:Bqhist} a histogram of the individual entries for the relationship given in Eq.~\ref{eq:powerlaw} is shown, for isotopes with  $Q > 2$ MeV, and with $g_A =1$.   The distribution shown in this figure is the basis for the numerical constant in Eq.~\ref{eq:powerlaw}.  Even though the individual priors for the matrix elements are not known, the central limit theorem evidently leads to a distribution that is approximately Gaussian.    Consistency with the central limit theorem over all isotopes and all theories lends  support to the conjecture that the various theoretical approaches are all calculating a single common observable $a_{0\nu}$, within the uncertainty expressed in the width, and the best estimate for that observable is the mean of the normal distribution.

  A reasonable estimate for the range of possible values for $g_A$  is $0.8\leq g_A\leq 1.269$.  The lower limit of the range is $0.8$ instead of $0.6$ in recognition of the correlation between the theory used to calculate the matrix elements and the extent of renormalization of $g_A$ required.  (This effect is responsible for the shape of the lower boundaries of the areas shown in Figs.~\ref{fig:DBDrate},\ref{fig:DBDrateAll}.)  If the prior for $g_A$ is assumed to be flat in $\log g_A$, the effect on the width of the distribution is relatively modest.  Including this contribution in quadrature  increases the standard deviation from 0.5 to 0.6:
\begin{eqnarray*}
a_{0\nu}&=& 10^{2.9\pm0.6} {\rm \ Mg}^{-1}{\rm \ y}^{-1}{\rm \ eV}^{-2}.
\end{eqnarray*}
In the future, if a more detailed understanding of the appropriate renormalization of $g_A$ for each theory emerges, there is the possibility that the width of the distribution could actually be reduced, perhaps quite substantially.  Simply using $g_A=0.8$ for the shell-model entries and 0.6 for QRPA and IBM would produce a narrower distribution when $g_A$ is incorporated with the matrix elements. 

It could be anticipated that some correlation between the energy release in a decay and the size of the matrix element might arise simply from differences in the nuclear structure between parent and progeny.  In the shell model, a large Q-value and high $Z$ both tend to accentuate differences in the neutron and proton valence shells involved in the transition.  However, because of the spin-orbit force, a monotonic trend is not in fact expected.  Specifically, $^{100}$Mo in the middle of the suite of isotopes is one of the few cases (perhaps the only case) where the spin-orbit partners are open in the ground states: the $\nu g_{7/2}$ valence neutrons can decay into $\pi g_{9/2}$ proton states.  A much larger space is needed for a good description, naturally, and there is unfortunately no large-basis shell model calculation for this nucleus.   Nevertheless, the QRPA, GCM, IBM, and PHFB models tend to give somewhat larger values for $^{100}$Mo, possibly reflecting this favorable situation.  In the IBM, the matrix elements display a parabolic dependence that peaks between closed shells and have a general dependence as $A^{-2/3}$.\cite{PhysRevC.79.044301}  A comprehensive understanding, across the theoretical spectrum, of the relationship captured in Eq.~\ref{eq:powerlaw} between the phase space and the matrix element  could provide useful additional insight into the predictive accuracy of the calculations of the nuclear matrix elements.

\section{Conclusions}

Large-scale experiments searching for neutrinoless double beta decay are in the planning and prototype  stages.  The importance of the objective has led to a  productive, intensive, and world-wide theoretical effort to make the best possible predictions for the nuclear matrix elements.  That effort continues apace.  

We have organized currently available theoretical input on nuclear matrix elements, phase-space factors, and $g_A$ in a way that is intended to be helpful to experimental design teams and to provide feedback to theoretical teams.  The results are somewhat surprising.  First, we find that there is little evidence on theoretical grounds to favor or disfavor the choice of one isotope over another.  All eleven for which a significant body of theoretical calculations exists seem to have about the same sensitivity to $0\nu\beta\beta$ decay per unit mass.  Second, we note a striking inverse correlation between the phase-space  available and the size of the nuclear matrix element.  There are essentially no exceptions, not even $^{48}$Ca, provided the Q-value is $> 2$ MeV.  Third, barring a large theoretical bias common to all existing theoretical methods, experiments at the tonne scale with negligible background can be expected to be sensitive to Majorana masses in the 10-100 meV range.   

The signal reported by Klapdor-Kleingrothaus and Krivosheina\cite{KlapdorKleingrothaus:2006ff} for $0\nu\beta\beta$ in $^{76}$Ge is 246(40) decays per tonne per year (1 tonne = 1 Mg = 1000 kg).  Upper limits have recently been reported for $0\nu\beta\beta$ in $^{136}$Xe by the EXO Collaboration\cite{Auger:2012ar} and the KamLAND-Zen Collaboration.\cite{Gando:2012zm}   The results are summarized in Table \ref{tab:rates}.
\begin{table}[h]
\tbl{Comparison of recent results on the rate for $0\nu\beta\beta$ decay.}
{\begin{tabular}{lcr} 
\toprule
Isotope \phantom{aaaa} &  Specific Activity  & \phantom{aaaa} Reference \\
& Mg$^{-1}$ y$^{-1}$  &   \\
\colrule
$^{76}$Ge & $246\pm 40$ & \refcite{KlapdorKleingrothaus:2006ff} \\
$^{136}$Xe & $<192$ (90\% CL) & \refcite{Auger:2012ar} \\
$^{136}$Xe & $<162$ (90\% CL) & \refcite{Gando:2012zm} \\
\multicolumn{3}{l} {Combined:} \\
$^{136}$Xe & $<90$ (90\% CL) &  \refcite{Gando:2012zm} \\
\botrule
\end{tabular}
\label{tab:rates}}
\end{table}
The most probable value for the difference between two samples drawn at random from a normal distribution is one standard deviation.  Within the framework of the empirical scaling rule Eq.~\ref{eq:powerlaw}  the difference in specific activity expected from the distribution in matrix elements is about a factor of 3.  (The uncertainty in the renormalization of $g_A$ is omitted for this purpose on the premise that the effective value of $g_A$ is likely to have much in common from isotope to isotope.)  As it happens, the experimental rate limit from $^{136}$Xe differs by about a factor of 3 from the positive result reported for $^{76}$Ge, a not improbable outcome in this analysis.  Although the $^{136}$Xe limit is a 90\% CL upper limit rather than a positive observation, the extent of disagreement is not substantial.   The difficulty in drawing conclusions from results in two different isotopes comes from the width of the distribution of specific activities (Fig.~\ref{fig:Bqhist}) arising from theoretical variance, which  is nearly normal in the logarithm.  A more detailed statistical analysis without using  the universality relation outlined here has been presented recently by Bergstrom,\cite{Bergstrom1212.4484} who reaches the conclusion that the Ge and Xe results are probably incompatible.  However, the widths of the individual nuclear matrix element distributions assumed are much narrower.       

As experiments searching for $0\nu\beta\beta$ advance toward very large scales, the  absolute specific activity appears to be largely independent of the choice of isotope.  Instead, the decision on which isotopes offer the best opportunities for scale-up may be based on more technical criteria: cost, redundant identification of candidate $0\nu\beta\beta$ events, background, resolution, and similar issues.    We conclude first that a theoretical understanding of the correlation between the nuclear matrix element and the effective value of the axial-vector coupling constant $g_A$ offers perhaps the most immediate path toward reducing the width of the distribution shown in Fig.~\ref{fig:Bqhist}.   Secondly, theoretical understanding of the inverse correlation between the phase space and the square of the nuclear matrix element will add confidence in the predictive accuracy of the theory of neutrinoless double beta decay.

\section*{Acknowledgments}
  
Discussions with F. Iachello are gratefully acknowledged.  We thank J.A. Detwiler and S.R. Elliott for reading the manuscript and for valuable comments.  The research has been supported by DOE through grant   DE-FG02-97ER41020.

\bibliographystyle{iopart-num}
\bibliography{DBDbib}

\end{document}